%% file: main-detection.tex
\documentclass[3p,review]{elsarticle}
\usepackage{amsfonts}
\usepackage{amsmath}
\usepackage{amssymb}
\usepackage{diagbox}
\usepackage{graphicx}
\usepackage{textcomp}
\usepackage{subfigure}
\usepackage{multirow}
\usepackage{algorithm, algpseudocode}
\usepackage{mathrsfs}
\usepackage{xcolor}
\usepackage{listings}
\usepackage{xspace}
\usepackage{enumitem}
\usepackage{pbox}
\usepackage{array}
\usepackage{ragged2e}
\usepackage{epstopdf}
\usepackage[colorlinks=true, linkcolor=blue, citecolor=blue]{hyperref}

\newtheorem{definition}{Definition}
\newcommand{\this}{\textit{Multi-IF}\xspace}
\newcommand{\avs}{ADVs\xspace}
\newcommand{\av}{ADV\xspace}

\newcommand{\numberthis}{\addtocounter{equation}{1}\tag{\theequation}}

\begin{document}
\sloppy

\begin{frontmatter}

\title{\this: An Approach to Anomaly Detection in Self-Driving Systems}

\author[buaa]{Kun Cheng}
\author[buaa]{Yuebin Bai}

\author[ntu]{Yuan Zhou\corref{cor1}}

\author[buaa]{Chao Yu}

\author[ntu]{Yang Liu}

\address[buaa]{School of Computer Science and Engineering, Beihang University, Beijing, China 100083}

\address[ntu]{School of Computer Science and Engineering, Nanyang Technological University, Singapore 639798}

\cortext[cor1]{Corresponding author}

\begin{abstract}
\input{abstract}
\end{abstract}


\end{frontmatter}

\input{chapter1}
\input{chapter2}
\input{chapter3}
\input{chapter4}
\input{chapter5}
\input{chapter6}

\bibliography{refs}

\end{document}

%% file: abstract.tex
Autonomous driving vehicles (ADVs) are implemented with rich software functions and equipped with many sensors, which in turn brings broad attack surface. 
Moreover, the execution environment of \avs is often open and complex.
Hence, \avs are always in risk of safety and security threats.
This paper proposes a fast method called \this, using multiple invocation features of system calls to detect anomalies in self-driving systems.
Since self-driving functions take most of the computation resources and upgrade frequently, \this is designed to work under such resource constraints and support frequent updates.
Given the collected sequences of system calls, the combination of different syntax patterns is used to analyze and construct feature vectors of those sequences.
By taking the feature vectors as inputs, one-class support vector machine is adopted to determine whether the current sequence of system calls is abnormal, which is trained with the feature vectors from the normal sequences.
The evaluations on both simulated and real data prove that the proposed method is effective in identifying the abnormal behavior after minutes of feature extraction and training.
Further comparisons with the existing methods on the ADFA-LD data set also validate that the proposed approach achieves a higher accuracy with less time overhead.

%% file: chapter1.tex

\section{Introduction}\label{intro}
Automobiles have been smarter than ever, resulting in the application of autonomous driving vehicles (\avs).
Current \avs are equipped with many intelligent devices, such as dozens of electrical control units (ECUs), a variety of sensors, and powerful computing platforms.

However, with the higher degree of autonomy, safety and security problems are escalated due to the complex software and the increased exposure of functionality to the adversaries.
On one hand, potential software bugs may lead to runtime errors, which put pedestrians, passengers, and vehicles at risk.
According to the open Autonomous Vehicle Disengagement Reports~\cite{dmv}, functional errors and system failures are the main causes of disengagement handling.
On the other hand, the broad attack surface makes intrusion possible.
For example, the vehicular networks, such as the vehicle to everything networks (V2X), Bluetooth, Wi-Fi hot spots, cellular networks, keyless entry systems, etc., make it possible for attackers to breach into the \av system.
Back in 2015, researchers hacked a running Jeep Cherokee by embedding a trojan program into one of its ECUs~\cite{greenberg2015hackers}.
In 2016, another group of researchers intruded a vehicle through a malicious application installed in the Android-based car-play system~\cite{ccs2016}, which could directly take control of the running car by fabricating control messages over the vehicular bus.
Keen Security Lab at Tencent successfully intruded a Tesla Model S remotely by exploiting a vulnerability of the embedded web browser in the central information display in 2016~\cite{keenlab2016}.
And they breached into a BMW i3 electric vehicle by compromising the telematics unit in 2018~\cite{keenlab2018}.

Anomaly detection becomes a vital task in an \av to guarantee its safe motion. 
Several mitigation measures have been proposed for some specific attacks, such as GPS spoofing and in-car communication fabrication.
However, as one of the most critical parts of an \av, there is little work on the mitigation of the self-driving software system.
The self-driving software system (often referred to as self-driving systems) of an \av conducts the software control logic of the vehicle, which performs most of the tasks during the motion, such as sensing the surrounding environment, planning the route, controlling the trajectory, etc.
More importantly, it is the only one generating control inputs to the \av's actuators/ECUs.
Thus, anomaly detection, as well as related protection, becomes more crucial and urgent to secure the self-driving system.

However, compared with general software systems, the anomaly detection for self-driving systems remains a challenging task for the following reasons.
\begin{itemize}[leftmargin=*]
\item Large-scale and complex software architecture.
Self-driving systems are usually massive, e.g., the open-source project \emph{Autoware}\footnote{\url{http://github.com/CPFL/Autoware}} contains over 270k lines of code, and Baidu \emph{Apollo}\footnote{\url{http://github.com/ApolloAuto/apollo}} has over 240k lines in its main functional modules currently.
Together with ECUs and other systems (e.g. car-play system), the software scale of an \av becomes tremendous.
The complexity of the software is also inherent to the massive functions of \avs.
To guarantee safe driver-less motion, a self-driving system needs to perform many tasks, such as object detection and tracking, localization, motion planning, and data fusion, each of which is a complicated task.
Their integration makes the whole system more complex.
Moreover, the self-driving system needs to continuously interact with actuators and the environment.

\item Complex and open environment. Usually, the execution environment of an \av, which is also the input space of the self-driving system, is with higher complexity.
On one hand, the environment is open and contains many unsafe factors, such as obstacles.
On the other hand, the environment is dynamic and partially unknown due to the existence of dynamic surrounding objects, such as other vehicles and pedestrians.
It is often challenging to accurately predict and track their occurrences/motion.

\item Non-deterministic behavior.
In a self-driving system, different (machine) learning algorithms/models are applied to complete some tasks, such as object detection.
Given the same data, different models may be obtained after the training phase.
The outputs of those models are challenging to predict, and the related components may react differently to the same input at different time instances.
Besides, due to the limitation of mechanisms' accuracy, a vehicle may perform the result of a self-driving system with some tolerant disturbance.
Hence, the motion of a vehicle, controlled by the self-driving system, may not be repeated exactly under the same (road) environment.
Due to the non-determinism, the validation of system behaviors is also challenging  (especially for the logic-/specification-based detection), as it is difficult to determine the normal disturbance of outputs between the vehicle and validation system, or the real output and specification.
\end{itemize}

In this paper, we propose a fast and ``inexpensive'' approach called \this, which utilizes \textbf{multi}ple \textbf{i}nvocation \textbf{f}eatures of system calls (syscalls) to detect anomalies in self-driving systems.
First, the sequence of syscalls is used to model the program behavior.
Syscall invocation is the essential programmatic way for user programs to directly request any privileged operations from the operating system kernel.
It has long been used in host-based intrusion detection systems, program profiling, malware analysis, etc.
Second, the syntax feature of syscall traces is extracted and analyzed to build an SVM based classifier.
Anomaly detection is essentially a classification problem.
The classifier should ideally learn from the normal behavior to set up the baseline, by which a pending trace can be examined whether it is compliant or not.
Thus, the one-class SVM is chosen due to its wide application in solving such issues.
At last, experiments are conducted on the real data from both our self-driving car and ADFA-LD~\cite{ADFA-LD} data set.
The evaluation shows that our approach achieves a high detection accuracy with minimal false alerts on the monitoring of self-driving functions.
Comparisons with existing works on the ADFA-LD data set proves that such an approach can achieve high detection accuracy and efficiency, and reduces the time overhead.

\this is inexpensive as 1) it combines cheap syntax features from testing data to achieve a more accurate detection engine; 2) the training is faster as it adopts an SVM-based solution (will be described in Section~\ref{training}). 
Note that with an inexpensive approach, it is necessary to not only work under the limitation of the onboard computer since most computation capability and resource should be allocated to perform  basic functionalities for safe motion, but also meet the requirement of rapid system/program update (e.g., Tesla \textit{Autopilot} updates up to hundreds of times for a build in one year~\cite{tesla-ota}), which also demands the detection engine to be upgraded quickly.

The contribution of this paper includes:
\begin{itemize}[leftmargin=*]
	\item We propose a syscall based method to model self-driving systems and prove it is effective, instead of looking into the complex software architecture and environment.

	\item By extracting multiple syntax features from syscall sequences, we show they can be used in a one-class SVM based classification to perform anomaly detection in self-driving systems.
	
	\item The proposed approach achieves better performance and requires low training costs than existing similar methods on the ADFA-LD data set.
\end{itemize}

The rest of this paper is organized as follows.
Section~\ref{related} states the related work.
Section~\ref{motivation} briefs preliminaries as well as the motivation of this work.
Section~\ref{method} presents the detail of our approach.
Section~\ref{eva} shows the evaluation results.
Finally, we conclude this paper in Section~\ref{conclusion}.
For the review purpose, all code and data could be found in \url{https://bitbucket.org/chengkunbuaa/avdetection_code/}.

%% file: chapter2.tex

\section{Related Work}\label{related}
\textbf{Attacks and protection in cyber-physical systems.}
Mitchell \emph{et al}~\cite{ai-uav} proposed an adaptive specification-based intrusion detection system (IDS) to detect malicious unmanned aerial vehicles (UAVs) in an airborne system.
Such an IDS monitored the output of embedded sensors and actuators, then defined behavior rules from their threat model.
Vuong \emph{et al}~\cite{rv-eval} developed a decision tree-based method to detect cyber attacks on a small robotic vehicle and tested it with both cyber and physical attacks.
The conclusion was that adding physical features could help improve detection accuracy.
Moosbrugger \emph{et al}~\cite{Moosbrugger2017} performed runtime monitoring on UAVs to detect threats such as the denial of service (DoS) or GPS spoofing by monitoring commands, signals, software behaviors, and so on.
Choi \emph{et al}~\cite{controlinvariants} proposed to use control invariants to detect external physical attacks by using a proportional-integral-derivative (PID) controller to examine whether the runtime behavior matches the controller.
However, those solutions require either pre-defined errors (e.g., decision tree solution) or detailed analysis of source code and binary executable files (e.g., control invariant modeling), which hinders the application in large systems.
Besides, F. Guo~~\emph{et al}~\cite{tvt-vad} used sensor data consistency and frequency to detect abnormal execution in autonomous driving networks.
K. Zhu~\emph{et al}~\cite{tvt-can} proposed an anomaly detection approach based on the long short-term memory (LSTM) network to check any CAN bus traffic from the time and data dimension, which resulted in a satisfactory accuracy.

Moreover, a variety of methods have been proposed to counter attacks against sensors or communication in vehicles.
For example, Park \emph{et al}~\cite{sensordtect} used pairwise inconsistencies between sensors to detect transient attacks or faults for GPS receivers.
Kar \emph{et al} \cite{ccs2014}~proposed an automated detection and vehicle identification system to mitigate GPS interference on the vehicle tracking system.
However, neither method was applicable as either multiple sensors or specific roadside units were required.
Cho and Shin~\cite{ccs2016} revealed a new type of DoS attack, which was caused by the vulnerability in in-vehicle networks and could disable ECUs via the error-handling mechanism.
However, the detection and mitigation required accurate time synchronization.
Bouard \emph{et al} \cite{bmw2013} proposed a decentralized information control approach to enhancing the security and privacy of in-car communication based on the deployed authentication framework for each ECU.
Woo \emph{et al} \cite{attackcan}
proposed an encryption and authentication protocol to protect the CAN bus.
However, the performance overhead of authentication in \cite{bmw2013, attackcan} limits their practical adoption.
Recently, Steger \emph{et al} proposed a framework for secure and dependable wireless software update on ECUs \cite{steger2018efficient}, which utilized \emph{IEEE} 802.11s mesh network and deployed a cryptography solution.
However, no internal computation platform security concerns such as system intrusion were considered.

\textbf{Syscall-based anomaly detection.}
Syscalls have long been used with anomaly detection or signature-based detection in host-based intrusion detection systems~\cite{syscall_ids}.
Anomaly detection often suffers from high the false-positive rate since it is difficult to establish a perfect baseline.
That is because the software execution is highly dynamic, and the complexity of modern computer systems makes it even harder to gather and process all normal execution data.
However, anomaly detection still plays an important role in defense as it assumes no prior knowledge of potential attacks, which greatly differs itself from signature-based detection.
Signature-based detection usually offers a low false alarm rate and high accuracy for the attacks that match the pre-collected data templates, but it cannot deal with unseen attacks.
Moreover, it relies on the accurate signature gathered from the attack evidence, which increases the application difficulty in a new system.

For syscall based intrusion detection, language modeling techniques are widely used \cite{syscall_ids,Bridges:2019}.
Among the recent anomaly detection works,
Marteau~\cite{tifs19} defined the covering similarity to measure a testing symbolic trace against a bunch of (normal) traces, which was used to identify the abnormal sequences.
However, since testing on ADFA-LD reached the best result when all 833 training traces and an additional 1000 traces from the validation set were used, such an approach acquired a rich feature set.
Besides, \cite{tifs19} tended to extract and build the optimal covering sets using all subsequences in the evaluation (though optimized algorithms were proposed).
By contrast, our approach reduces the pattern and feature sets by adopting the proposed 3-step method in Section~\ref{training}, which cuts down the storage and training overhead. 
Khreich \emph{et al}~\cite{KHREICH2018415} combined different detection methods, namely the Sequence Time-Delay Embedding (STIDE), Hidden Markov Model (HMM), and one-class SVM to improve the accuracy.
Although computing the combination of different detectors was fast, training all detectors was computing-intensive and time-consuming.
Creech \emph{et al} \cite{creech} used context-free semantic features in syscall traces, together with an extreme learning machine, to build a neural network based classifier.
They achieved almost the perfect performance on the KDD98 data set, but the approach was computational heavy, which took weeks to extract the semantic features and days for training.
With the semantic features and SVM, they also achieved a good result on the ADFA-LD data set.
Xie \emph{et al} \cite{mxie1-1,mxie1-2}  used syntax features, such as the length of a syscall trace and the relative frequency of individual calls, and the k-NN and k-mean clustering models to achieve acceptable results.
In another work~\cite{mxie2}, they used short sequences and frequencies to train a one-class SVM classifier, which improved the accuracy.
Haider \emph{et al} \cite{whaider} used four statistical features in a trace, i.e., the least/most repeated and the minimum/maximum values, to detect attacks.
Three learning algorithms were applied to improve the performance over Xie's works while achieving a fast training, including SVM with linear kernels, SVM with radial basis kernels and k-NN.
Huorong \emph{et al.}~\cite{REN201752} segmented sequence data by a sliding window to build a dynamic Markov model to analyze their simulated data set and an airport traffic data set, which improved the adaptability and stability when compared with the classical Markov approaches.
On the UNM data set, Hoang \emph{et al}~\cite{HOANG20091219} proposed to use the Hidden Markov model to examine normal syscall sequences and generate four pattern sets.
Khreich~\emph{et al}~\cite{KHREICH2017186} used various n-grams and their frequencies as features, which was exactly the complete large pattern set (all L-$k$ clusters) extracted in Section~\ref{patterns}.
However, it would lead to enormous feature vector space (e.g. there were 142,190 features when N=6). 
And as the average Euclidean distance was used to determine the similarity of a testing trace to all normal sequences, it would involve a large amount of computation.
Fuzzy rules are applied to check whether the tested trace is normal or not by considering the produced probability and pattern frequencies.
Results showed they reduced the false alarm rate by almost a half.
Although the above studies have made remarkable contributions to the host-based intrusion detection system (HIDS), they are either time/resource-consuming or less accurate, which are too imperfect to be used in \avs .
Thus, a faster and more accurate detection method is required for the current defense system for \av. 

%% file: chapter3.tex

\section{Preliminaries and Motivation}\label{motivation}

\subsection{The Self-Protection Framework for \avs}\label{guardauto}
Robot Operating System (ROS), built on the top of the Linux kernel, is an open-source and flexible framework for developing robot control systems, which is prevailing in robotics.
Currently, we have been working on a self-protection framework with great flexibility and extensibility for ROS-based self-driving systems~\cite{cheng2020guardauto}.
In this framework, hardware-assisted virtualization was used to isolate different software components of a self-driving system.
Each isolated software component, referred to as a partition, is equipped with a self-protection subsystem to inspect the partition execution and plan mitigation measures.

As an important part of the efforts made to secure the cyber world of autonomous driving platforms.
The anomaly detection is designed to handle those possible threats from both inside and outside, such as malicious intrusion (e.g., a compromised partition image during cloud update) or runtime faults in a partition (may lead to system failure/malfunctioning).
Together with the efforts made in securing the physical world (the motion-based detection), we hope to build a complete model to better explain the overall status of the vehicle, and to locate the possible root cause of any detected abnormal functioning of the whole CPS system.

To secure the cyber world, anomaly detection becomes vital.
The main steps to detect abnormal execution are selecting proper information to monitor and proposing a proper analysis method to check whether the monitored information is correct.
Due to the complexity of architecture and program/code logic in self-driving systems, monitoring and analyzing the execution of each partition directly may be challenging.
Critical execution paths may serve the monitoring well, but it is also important to remain stealthy and do not alert the intruder.
Thus, code instrument solutions are not suitable for our purpose.
Inspired by the virtual machine introspection (VMI) based techniques~\cite{xenaccess, drakvuf}, the sequence of syscalls invoked by a partition should be a good choice to develop new methods to extend and improve the current MAPE loop.

\subsection{Syscall and Anomaly Detection}
Modern Linux kernel provides over 300 syscalls \cite{syscall}, which are used by running processes to interact with the operating system kernel.
On one hand, the usage of syscalls is determined by the program source code and the libraries it relies on;
on the other hand, the invocation sequence of syscalls is highly dynamic and determined by the program logic and input data.
Hence, syscall sequences are often used for (host-based) intrusion detection in systems~\cite{Bridges:2019}, under the assumption that only running processes can harm the system and any damage can happen only through privileged operations.
Language models are prevalent for syscall-based intrusion detection, where sequential features, such as n-grams, are often used to build various detection engines.

Anomaly detection plays an important role in defense as it does not assume the prior knowledge of potential attacks.
It usually works by first establishing a baseline for the behavior of the  protected targets.
Ideally, if such a baseline is sufficiently accurate, any anomaly will be identified as a real threat.
But establishing such a perfect baseline is difficult and sometimes impractical, since 1) the software execution environment is highly open,  dynamic, and complex, and 2) the complexity of modern software systems makes it even harder to gather and process all normal execution data.
It is often required to upgrade the detection engine regularly with more evidence (training data) either online (e.g. semi-supervised learning) or offline (e.g. retraining).
However, the online upgrade is challenging as it is difficult to distinguish a rare-seen normal trace from real threats.
Thus, we try to accelerate the offline upgrade by reducing the training overhead, which is also what ``inexpensive'' here stands for.

\subsection{Support Vector Machine}\label{ocsvms}
Support Vector Machine (SVM) is a supervised machine learning model for classification and regression.
The basic idea of SVM for linear classification is to determine optimal hyperplanes to maximize the margin between any different groups of data.
In case they cannot be separated linearly, SVM uses kernel functions to map the data into a higher-dimensional space such that they can be separated in the new space.
One-class SVM is essentially a regular binary-class SVM where all training data belongs to the same class.
There are two popular kinds of one-class SVMs: one-class SVM according to Sch{\"o}lkopf, where the boundary is a hyperplane \cite{scholkopf2000support}, and another one according to Tax and Duin, where the boundary is a sphere \cite{tax2004support}.
One-class SVM is popular for anomaly detection as we conclude in Section~\ref{related}.

In a self-driving platform such as mentioned in Sec.~\ref{guardauto}, GPU is exclusively used by the object detection function (in our platform it is done via device passthrough). 
Moreover, it is tedious to maintain and update traditional intrusion detection systems installed on every host, which requires approaches that can reduce the time overhead while keeping the acceptable detection accuracy~\cite{liu2019host}.
As it is critical to develop a lightweight detection engine that would not potentially consume too many resources or introduce too much overhead, SVM-based approaches are preferred in this work.

%% file: chapter4.tex
\section{Methodology}\label{method}
The detection goal here is to identify the abnormal execution trace from the normal ones.
As described in \cite{creech}, given the set of valid subsequences, called patterns, extracted from normal traces, the occurrence of these subsequences in a new normal trace is significantly greater than those in an abnormal one.
Hence, the appearance of syscall patterns in a pending trace is vital to detect anomalies.
There are two main steps for such pattern-based detection.
The first one is pattern extraction.
In this work, the subsequences of contiguous syscalls are used as patterns.
How to find a proper set of subsequences with different lengths is the first important thing for accurate prediction.
The second one is building the detection engine .
As abnormal behavior is rare but catastrophic in self-driving systems, the training data is collected from the clean and safe execution of the target system.
Hence, how to use normal data to get a classifier to detect abnormal data is another challenge.
In this section, details of the proposed method will be presented.

\subsection{The Framework of the Proposed Method}
Let $\Sigma$ be the set of all allowed syscalls by the kernel.
Given the kernel source code, $\Sigma$ is a finite and deterministic set.
For example, as defined in \emph{/usr/include/x86\_64-linux-gnu/asm/unistd\_64.h}, Ubuntu kernel 4.4.108 x64 contains 326 syscalls, such as `0' for `\emph{read}', `1' for `\emph{write}' and `96' for `\emph{gettimeofday}'.
An invocation trace, denoted as $\pi$, is a time-ordered sequence of syscalls, i.e., $\pi = \{\xi_1, \xi_2, \ldots\}$, where $\pi[i]=\xi_i \in \Sigma$.

\begin{figure}
	\centering
	\includegraphics[width=0.4\linewidth]{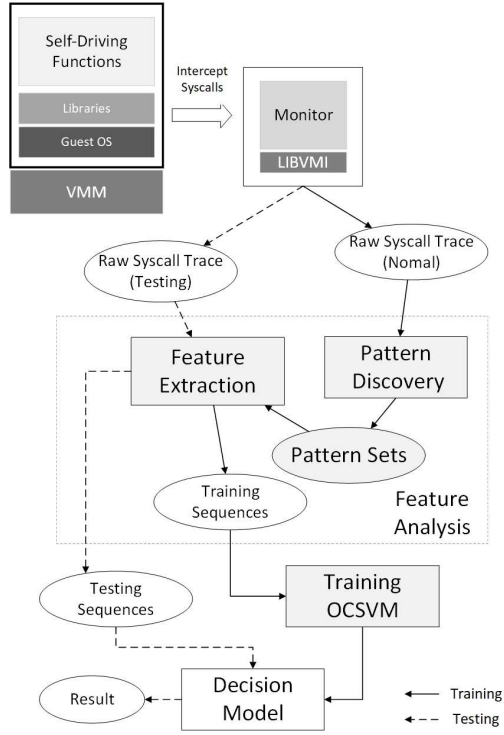}
	\caption{Anomaly detection procedure.}
	\label{flow}
\end{figure}

Fig.~\ref{flow} shows the whole structure of the proposed method.
Suppose $S_t$ is a set of normal syscall traces, and $S$ is the testing set containing both normal and abnormal syscall traces.
Our method includes two steps.
The first one is pattern discovery and feature extraction from $S_t$.
This step performs the preprocessing of the raw syscall sequences in $S_t$ and generates the inputs for decision making.
We first extract and build the set of patterns from the normal data.
Then we classify those patterns into different clusters based on their length, and calculate their frequencies in a trace as the features of this trace.
The second step is to train a detection engine.
Since the amount of normal data is usually larger than that of abnormal data, we adopt one-class SVM to construct the decision engine.
The input to the detection engine is the extracted feature vectors of syscall traces.

\subsection{Feature Extraction}\label{patterns}
Before diving into details, some basic definitions applied in this paper shall be given.
Based on the $n$-gram model, patterns can be defined by Definition \ref{defn_pattern}.

\begin{definition}[Pattern]\label{defn_pattern}
	Given a set of normal syscall traces $S_t = \{\pi_t: \forall t\}$, a syscall pattern is a $k$-gram in  $S_t$, denoted as $\pi_j^k$, where $1\leq k \leq \max_{t} |\pi_t|$.
	The set of all patterns is denoted as $\Psi=\cup_{k}\cup_j \{\pi_j^k\}$.
\end{definition}

The scale of patterns extracted from normal traces may be very large.
The direct application of these patterns may cause high dimensions of the input data for training and detection, which could result in unnecessary overhead.
Hence, further clustering is applied to reduce the number of used patterns.

\begin{definition} [L-$k$ Cluster]\label{defn_cluster}
	Given the pattern set $\Psi$, its L-$k$ cluster, denoted as $\Gamma_k$, is the set of all $k$-grams in $\Psi$, i.e., $\Gamma_k=\cup_{j: \pi_j^k\in \Psi} \{\pi_j^k\}$. All clusters in $\Psi$ are denoted as $\Phi=\{\Gamma_{k_1}, \Gamma_{k_2}, \ldots, \Gamma_{k_m}\}$.
\end{definition}

Based on this definition, all patterns in a cluster $\Gamma_k$ have the same sequence length $k$.

\begin{definition}[Feature]\label{defn_feature}
	Given the clusters $\{\Gamma_{k_1}$, $\Gamma_{k_2}$, $\ldots$, $\Gamma_{k_m}\}$, the feature of $\Gamma_k$ is defined as the frequency of $\Gamma_k$ in a trace $\pi$, denoted as $f_{\Gamma_k}(\pi)$. $f_{\Gamma_k}(\pi)=\sum_{\pi_j^k\in \Gamma_k} f_{\pi_j^k}(\pi)/|\pi|$, where $f_{\pi_j^k}(\pi)$ is the frequency of $\pi_j^k$ in $\pi$, and $|\pi|$ is the number of syscalls in $\pi$.
\end{definition}

For example, given a set of normal syscall traces, suppose $\Gamma_3$ $=$ $\{(22, 1, 1)$, $(0, 22, 23)$, $(1, 96, 1)$, $(96, 5, 128)\}$ and $\Gamma_5$ $=$ $\{(0, 22, 23 ,1, 5)$, $(96 , 5, 128, 4, 34)$, $(1, 5, 96 , 5, 1)\}$.
If there exists a trace $\pi$ $=$ $\{0$, $22$, $23$, $1$, $5$, $96$, $5$, $128$, $4$, $34\}$, then we have $f_{\Gamma_3}(\pi) = 2/10 = 0.2, f_{\Gamma_5}(\pi) = 2/10 = 0.2$.

\textbf{Pattern cluster building}.
The first thing for feature extraction is to build the set of $L$-k clusters.
Given a value $k$, the construction of $\Gamma_k$ can be described by function BUILD\_SET($S_t,k$) (Lines \ref{alg_s1}$-$\ref{alg_e1}) in Algorithm \ref{alg:cs}, where Lines \ref{alg_t1}$-$\ref{alg_t2} show the extraction of $k$-grams in each trace $\pi$.
More specifically, the algorithm first searches for all $k$-grams in one trace from the first $k$ contiguous syscalls to the last $k$ contiguous syscalls.
Hence, there are total $|\pi|-k+1$ iterations of Lines \ref{alg_t3}$-$\ref{alg_t2} for a trace $\pi$.
Once it completes on the current trace, the algorithm switches to check the next one, until all traces in $S_t$ are processed.

\begin{algorithm}[!t]
	\footnotesize
	\algtext*{EndWhile}
	\algtext*{EndFor}
	\algtext*{EndIf}
	\algtext*{EndFunction}
	\caption{Feature extraction for L-$k$ cluster.}
	\label{alg:cs}
	\algnewcommand{\LeftComment}[1]{\Statex \(\triangleright\) #1}
	\renewcommand{\algorithmicrequire}{\textbf{Input:}}
	\renewcommand{\algorithmicensure}{\textbf{Output:}}
	\begin{algorithmic}[1]
		\Require

		\Statex $k$: an integer denotes the length
		\Statex $S_t$: the training data set
		\Statex $S$: the testing data set
		
		\Statex 
		
		\Function{build\_set}{$S_t$, $k$} \label{alg_s1}
			\State $\Gamma_k \leftarrow \emptyset$
			\For { each $\pi \in S_t$} \label{alg_t1}
				\For {$i=1: |\pi|-k+1$} \label{alg_t3} 
					\If{$\pi[i,i+k-1]$ not in $\Gamma_k$}
						\LeftComment $\pi[i,i+k-1]$ is the subsequence of syscalls from the $i$-th syscall to $(i+k-1)$-th syscall in $\pi$.						
						\State add $\pi[i,i+k-1]$ into $\Gamma_k$ \label{alg_t2}
					\EndIf
				\EndFor
			\EndFor		
			\State \Return $\Gamma_k$
		\EndFunction \label{alg_e1}
		\Statex
		
		\LeftComment{function to compute the frequency of $\Gamma_k$ in a given trace $\pi$}
		\Function{eval\_trace}{$\pi$, $\Gamma_k$}\label{alg_f2s}
				\State $f_{\Gamma_k}(\pi) \leftarrow 0$
				\For {each $\pi_j^k \in \Gamma_k$}
					\State $f(\pi_j^k)$ $=$ the times that $\pi_j^k$ appears in $\pi$; \label{alg_f2_1}
					\State $f_{\Gamma_k}(\pi) \leftarrow f_{\Gamma_k}(\pi) + f(\pi_j^k)$; \label{alg_f2_2}
				\EndFor
				\State \Return $f_{\Gamma_k}(\pi) \leftarrow f_{\Gamma_k}(\pi)/|\pi|$; \label{alg_f2_3}
		\EndFunction \label{alg_f2e}
	\end{algorithmic}
\end{algorithm}

Take a normal trace $\pi=\{\xi_1, \xi_2, \xi_3, \xi_4, \xi_1, \xi_2, \xi_3\}$ as an example to illustrate the performed Line \ref{alg_t3}$-$\ref{alg_t2}.
Suppose $k=3$ and currently $\Gamma_3=\emptyset$.
Since $|\pi|=7$, Lines \ref{alg_t3}$-$\ref{alg_t2} will be iterated 5 times to generate $\Gamma_3$ from $\pi$.
For the first iteration, the 3-gram is $\xi_1\xi_2\xi_3 \triangleq \pi_1^3$.
Since $\Gamma_3=\emptyset$, $\pi_1^3$ is added to $\Gamma_3$, resulting in $\Gamma_3=\{\pi_1^3\}$.
The next 3-gram is $\xi_2\xi_3\xi_4 \triangleq \pi_2^3$.
Since $\pi_2^3$ is not in $\Gamma_3$, $\Gamma_3=\{\pi_1^3, \pi_2^3\}$.
Similarly, $\xi_3\xi_4\xi_1 \triangleq \pi_3^3$ and $\xi_4\xi_1\xi_2 \triangleq \pi_4^3$ are added to $\Gamma_3$ sequentially, generating $\Gamma_3=\{\pi_1^3, \pi_2^3,\pi_3^3,\pi_4^3\}$.
The last iteration checks the final 3-gram in $\pi$, i.e., $\xi_1\xi_2\xi_3$.
As it has been in $\Gamma_3$, there is no need to add it again.

\textbf{Feature extraction}.
In this paper, the frequency of each cluster is used as features to characterize the property of a trace.
Based on Definition \ref{defn_feature}, given a trace $\pi$, the frequency of $\Gamma_k$ can be computed by function EVAL\_TRACE($\pi$, $\Gamma_k$) (Lines \ref{alg_f2s}$-$\ref{alg_f2e}) in Algorithm \ref{alg:cs}.
For each pattern $\pi_j^k$ in $\Gamma_k$, the function first counts the number of its occurrence in $\pi$ (Line \ref{alg_f2_1}), and then updates $f_{\Gamma_k}(\pi)$ (Line \ref{alg_f2_2}).
Since different traces may have different lengths, the absolute amounts of occurrence of $\Gamma_k$ in different traces may be significantly different, which can cause bias during training and prediction.
Hence, we normalize the occurrence amount of $\Gamma$ in a trace by dividing the trace length, i.e., Line \ref{alg_f2_3} in Algorithm \ref{alg:cs}.

By checking all $\Gamma_k$ in $\Phi$, we can count their frequencies in a trace.
Hence, given an arbitrary $\pi$, we can measure it via the following $m$-dimensional vector.
\begin{equation}\label{tranform}
\footnotesize
	f(\pi) = (f_{\Gamma_{k_{1}}}(\pi), f_{\Gamma_{k_2}}(\pi),..., f_{\Gamma_{k_{m}}}(\pi))^T
\end{equation}
where $m=\|\Phi\|$ and $f_{\Gamma_{k_{i}}}(\pi)$ is computed from EVAL\_TRACE($\pi$, $\Gamma_{k_i}$).
Clearly, with \eqref{tranform}, we can translate the trace space to a subset of $\mathcal{R}^m$, where $\mathcal{R}^m$ is the m-dimensional Euclidean space.
Suppose the translated spaces of $S_t$ and $S$ are $\Omega_t$ and $\Omega$, respectively.

\subsection{One-Class SVM based model training}\label{training}
In this paper, we use a popular tool called LIBSVM\cite{libsvm}.
As pointed out by~\cite{khan2014one}, the method of Sch{\"o}lkopf et al.~\cite{scholkopf2000support} and the SVDD method of Tax and Duin~\cite{tax2004support} operate comparably and both perform best when the Gaussian kernel is used in practical implementations, we choose one-class SVM (OCSVM) with (Gaussian) radial basis function (RBF) 
to train the prediction model with the normal data.
Such one-class SVMs applied here is according to Sch{\"o}lkopf's work.
By walking through the SVM model, we clearly explain what parameters are required and how they are determined.

In LIBSVM\cite{libsvm}, training an OCSVM model with RBF kernel is to solve \eqref{finalsvm}.
\begin{align*}\label{finalsvm}
\min_{\alpha}\ &\frac{1}{2} \sum_{i,j=1}^{n}\alpha_i\alpha_j\exp(-\mu||x_i - x_j||^2) \\
\text{s.t. }\ & 0\leq \alpha_i \leq 1/(n\nu) \numberthis \\
& e^{T}\alpha = n\nu
\end{align*}
where $\alpha_i, i=1,2,\ldots,n$, are Lagrange multipliers or dual variables, and $\alpha=(\alpha_1,\alpha_2,\ldots,\alpha_n)^T$; $e = [1, . . . , 1]^T$
is the vector of all ones, $\mu = 1/||x_i||$ by default.
Thus, $\nu$ is the only one required to be tuned.
Because different $\nu$ may generate various optimization problems of \eqref{finalsvm}, which in turn affect the solution of $\alpha_i$.
Here, We use the grid search method to determine $\nu$ from the set \{0.5 (default), 0.2, 0.1, 0.005, 0.001, 0.0005, 0.0001\}, which is collected in early attempts.

\begin{figure}[t!]
	\centering
	\includegraphics[width=0.5\linewidth]{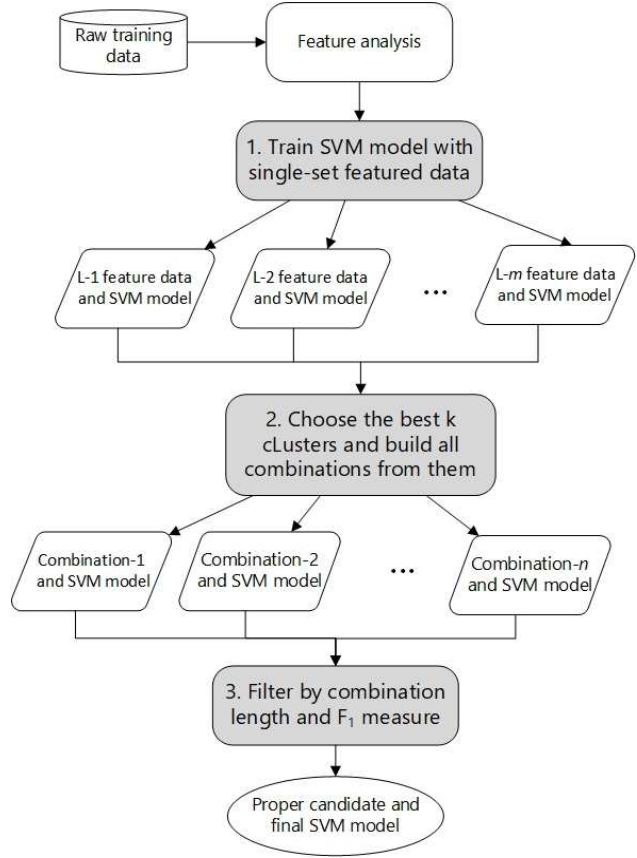}
	\caption{Procedure of training the detection model.}
	\label{method:training}
	\vspace{-10pt}
\end{figure}

\begin{table*}[!t]
	\footnotesize
	\caption{ADFA-LD early results on different clusters (step 1) [\%].}
	\begin{center}
		\begin{tabular}{|c|c|c|c|c|c|c|c|c|c|c|c|c|c|c|c|c|c|c|}
			\hline
			\multirow{2}{*}{\diagbox{$k$}{$\nu$}} &  \multicolumn{ 2}{c|}{0.5} & \multicolumn{ 2}{c|}{0.2} & \multicolumn{ 2}{c|}{0.1} & \multicolumn{ 2}{c|}{0.05} & \multicolumn{ 2}{c|}{0.01} \\ \cline{ 2- 11}
			\multicolumn{ 1}{|c|}{} & FAR & DR & FAR & DR & FAR & DR & FAR & DR & FAR & DR \\ \hline
			1 & 30.695 & 77.748 & 12.557 & 28.150 & 5.993 & 11.528 & 5.672 & 10.456 & 4.414 & 8.847 \\ \hline
			2 & 65.027 & 66.890 & 8.669 & 7.775 & 4.071 & 4.155 & 2.653 & 4.558 & 1.212 & 3.619 \\ \hline
			3 & 40.096 & 78.954 & 15.256 & 41.957 & 6.725 & 33.780 & 5.833 & 33.110 & 4.140 & 30.831 \\ \hline
			4 & 42.864 & 92.627 & 27.196 & 67.024 & 21.523 & 62.064 & 19.511 & 61.260 & 17.383 & 60.188 \\ \hline
			5 & 72.210 & 98.257 & 34.149 & 93.566 & 31.084 & 89.678 & 28.454 & 87.265 & 26.189 & 84.316 \\ \hline
			6 & 74.909 & 99.330 & 40.393 & 98.794 & 36.002 & 95.040 & 34.218 & 92.895 & 31.999 & 92.493 \\ \hline
			7 & 77.493 & 99.598 & 43.207 & 98.928 & 38.701 & 95.174 & 36.848 & 93.164 & 35.522 & 93.164 \\ \hline
			8 & 80.764 & 99.866 & 48.605 & 98.525 & 42.978 & 95.040 & 40.714 & 93.164 & 38.426 & 93.164 \\ \hline
			9 & 60.476 & 99.732 & 53.088 & 96.381 & 48.399 & 94.370 & 45.334 & 93.700 & 43.435 & 93.700 \\ \hline
			10 & 61.528 & 99.732 & 53.728 & 96.381 & 50.114 & 94.504 & 46.935 & 93.834 & 44.831 & 93.834 \\ \hline
			11 & 63.701 & 99.464 & 54.140 & 95.576 & 51.189 & 94.102 & 49.062 & 93.834 & 47.027 & 93.834 \\ \hline
			12 & 64.478 & 97.185 & 54.872 & 94.906 & 52.196 & 93.834 & 50.160 & 93.834 & 47.850 & 93.834 \\ \hline
			13 & 65.599 & 97.051 & 56.176 & 94.504 & 52.722 & 93.834 & 50.595 & 93.834 & 48.673 & 93.700 \\ \hline
			14 & 65.759 & 98.257 & 58.326 & 94.906 & 55.627 & 94.102 & 51.715 & 93.968 & 49.611 & 93.700 \\ \hline
			15 & 67.109 & 98.257 & 58.875 & 96.113 & 55.855 & 94.102 & 53.317 & 93.834 & 50.069 & 93.834 \\ \hline
		\end{tabular}
	\end{center}
	\vspace{-10pt}
	\label{adfa-single}
\end{table*}

In this paper, each input $x_i$ for the OCSVM is a vector in $\Omega_t$ or $\Omega$.
Specifically, the data in $\Omega_t$ is used to train the SVM model, and that in $\Omega$ is used to test the trained model.
For each trace $\pi_i$, its feature vector is $x_i=(f_1,f_2,\ldots,f_m)^T$. If $\pi_i$ is a normal trace, then $x_i$ is labeled as $+1$, otherwise labeled as $-1$.
Thus, given  $x_i$ and its label, the input format for LIBSVM can be written as  $\left\lbrace label\quad 1: f_1\quad 2: f_2\quad ...\quad m: f_m \right\rbrace$, where $label= +1$  or  $-1$.

Given a set of normal syscall traces, the simplest way to determine the clusters is to select all possible lengths, from $1$ to the length of the longest trace.
However, the number of clusters may also become very large.
Hence, to optimize the training cost and detection effectiveness, we need to deal with the following issues in practice.
\begin{itemize}
	\item The dimension of input data. The number of clusters to be selected for feature extraction has to be decided properly since it determines the dimension of data space of SVM, which affects the training cost significantly.
	
	\item The combination of different clusters. The feature vector is based on the cluster set $\Phi$. Different combinations of clusters will result in various training data, which directly affect the performance of classification. Thus, a proper set of clusters has to be determined.
\end{itemize}

To solve the above issues, a 3-step method is proposed to choose the clusters and get a proper combination to train a detection model.
The whole procedure is shown in Fig.~\ref{method:training}.
In the sequel, we take the ADFA Linux data set (ADFA-LD) \cite{ADFA-LD} as an illustrative example to show a heuristic process for cluster selection.

\subsubsection{Probing potential clusters}
To find out a proper set of clusters, the first step is to test the classification performance by taking the features of a single cluster (as shown in Fig.~\ref{method:training}, whose length varies from 1 to 20, as inputs and checking with different $\nu$, $\nu\in \{0.5,0.2,0.1,0.05,0.01\}$.
The results are shown in Table~\ref{adfa-single}.
As our goal is to distinguish abnormal traces from normal ones, as well as to easily compare the final result with those published by similar researches~\cite{creech,tifs19,mxie1-1, mxie1-2,mxie2, whaider}, the detection focuses on the false alarm rate (FAR) and detection rate (DR), whose definitions are given in \eqref{far},
\begin{align}
	\footnotesize
	DR &= \frac{\text{\small \# of Correctly Detected Abnormal Traces}}{\text{\small \# of Abnormal Traces }}\nonumber\\
	   &= \frac{TP}{TP+FN} \nonumber \\ 
	\nonumber\\
	FAR &= \frac{\text{\small \# of Wrongly Detected Abnormal Traces}}{\text{\small \# of Normal Traces}} \nonumber\\
	    &= \frac{FP}{FP+TN} \label{far}
\end{align}
where $TP$, $TN$, $FP$ and $FN$ stand for the numbers of true positives, true negatives, false positives, and false negatives, respectively, regarding an abnormal trace as a positive result, as shown in Table \ref{measure-tab}.

\begin{table}[t!]
\centering
\caption{Meanings of $TP$, $TN$, $FP$ and $FN$.}
\footnotesize
\begin{tabular}{|c|c|c||c|}
\hline
    \backslashbox{Actual}{Detected} & \begin{tabular}[c]{@{}c@{}}Abnormal\\ (Positive)\end{tabular} & \begin{tabular}[c]{@{}c@{}}Normal\\ (Negative)\end{tabular} & Total \\ \hline
Abnormal & $TP$  & $FN$  &  $N_2$    \\ \hline
Normal   & $FP$  & $TN$  &  $N_1$    \\ \hline
\end{tabular}
\label{measure-tab}
\vspace{-10pt}
\end{table}

\subsubsection{Determining the optimal clusters}\label{three-steps}
The second step is to choose the candidate set and test each possible combination, as shown in Fig.~\ref{method:training}.
The key here is to decide the maximum pattern length $K_{max}$ based on the results from the first step.
The goal of increasing $k$ is to achieve better performance.
On one hand, increasing $k$ could eventually yield more feature data, which generally benefits the training, if it can increase DR and suppress (decrease or at least no to increase quickly) FAR.
On the other hand, $k$ should be capped to lower the storage and computation overhead once the increment of $k$ does little help in differentiating DR and FAR.

From the observation of early results presented in step-1 (e.g., Table~\ref{adfa-single}), it is a general trend that during the given range of $k$ probed in the first step: for each given $\nu$, 1) both DR and FAR increase with the growth of $k$ before they reach the utmost; 2) the increment becomes slow when they are reaching the utmost. 
In such a situation, $K_{max}$ can be decided by analyzing the change rate ($\Delta$) of DR and FAR, which can be described as \eqref{delta_rate}.

\begin{align}\label{delta_rate}
\footnotesize
\Delta^{DR}_{k, \nu} & = DR_{k,\nu} - DR_{k-1,\nu}, \\
\Delta^{FAR}_{k, \nu} & = FAR_{k,\nu} - FAR_{k-1,\nu},\ k=2,3,4,...,N\nonumber
\end{align}
where $N$ is the range of $k$ probed in the first step.
Thus, $K_{max}$ can be decided by \eqref{max_k}.
\begin{align}\label{max_k}
\footnotesize
K_{max} &= \arg \min_{k}\Delta_{k, \nu}  \\
\textit{s.t } \Delta_{k, \nu} = &\Delta^{DR}_{k, \nu} - \Delta^{FAR}_{k, \nu} < 0,
 k=2,3,\ldots,N. \nonumber
\end{align}
According to \eqref{max_k}, $k = K_{max}$ is the very point where $\Delta_{k, \nu}$ decreases most, which means that keeping increasing k (while $k > K_{max}$) helps little in differentiating DR and FAR in the given range of k presented in the first step.

Back to the example, as shown in Table~\ref{adfa-single},  FAR and DR almost stop increasing (or we could say the increment is little) when $k > 9$.
Thus, the clusters from L-1 to L-9 can form a candidate set for potential feature extraction, as they yield better detection performance.  
When searching for a proper set of clusters, possible combinations can range from 2 clusters to 9 clusters in this case.
Indeed, for $m$ selected clusters, there are $\sum_{j=2}^m \mathrm{C}_m^j = 2^{m} - (m+1)$ combinations.
Since the computation of feature (i.e., cluster's frequency) is independent,
evaluating a trace with multiple clusters can be directly done by collecting the frequency of every single cluster from the step-1, and concatenating them to get new input data for the SVM model.
For example, suppose the frequencies of L-3, L-4 and L-7 clusters are $f_1$, $f_2$ and $f_3$ in a given trace individually, we can get a new 3-dimensional data consisting of those clusters to describe this trace, by directly concatenating them:
\begin{equation*}
\footnotesize
	label\quad 1: f_1 \quad 2:f_2\quad 3:f_3
\end{equation*}
As reducing operation takes $O(1)$ time complexity, we can try different combinations with brute force.

Given the fact that training and evaluating an SVM takes only seconds as shown in our evaluation, trying all possibilities depends on the number of cluster combinations.
Thus, the training with multiple sets takes $O(2^m)$ time complexity, where $m = K_{max}$.
Back to the example, When such an approach was applied to ADFA-LD, it took minutes (wall time) to finish the training process, from feature extraction to getting all multi-set evaluation results.

\subsubsection{Deriving the best cluster combination}
However, different combinations may perform similarly.
In such a case, two rules are proposed to choose the best one, which is also the final step to achieve the detection model as shown in Fig.~\ref{method:training}.
First, when multiple candidate combinations are available, the one with the shortest combination length is the best.
Here, the \textit{combination length} is defined as the sum of the length of all patterns used in the combination.
For example, for \{L-3, L-4, L-7\}, the combination length is $3+4+7=14$.
Then, if there are still more than one candidate, we use  $F_1$ measure to choose the best one.
$F_1$ measure is the harmonic average of the precision and recall, which is defined in ~\eqref{fm}.

\begin{align}
\footnotesize
F_1 &= 2 \cdot \frac{Precision \cdot Recall}{Precison + Recall} \cdot 100\% \nonumber\\
&= \frac{2}{1+\frac{1}{DR}+\frac{FAR}{DR}\cdot\frac{N_1}{N_2}} \cdot 100\% \numberthis \label{fm}
\end{align}
where
\begin{align}
\footnotesize
	Recall &= DR \nonumber\\	
	Precision &= \frac{\text{\small \# of Correctly Detected Abnormal Traces}}{\text{\small \# of Detected Abnormal Traces}} \nonumber\\
	&= \frac{TP}{TP+FP}  \nonumber
\end{align}
where $N_1$ is the number of normal testing traces, and $N_2$ is the number of abnormal testing sequences.
In such a case, the one with the largest $F_1$ will be considered the best.

%% file: chapter5.tex

\section{Evaluations}\label{eva}
\subsection{Experiment Setups}
The testbed was equipped with Intel Xeon E5-1650 v3 CPU, 32GB Memory and 1TB disk. The host operating system was Ubuntu 16.04.3 x64, with Linux kernel version 4.4.108. 
We isolated either of localization and mapping components in a Xen virtual machine (VM) with 2 virtual CPUs and 2GB memory and deployed the VMI-based monitor to capture the execution trace for analysis.

The evaluation contains two parts.
In the first one~(Sec.~\ref{eval_gnss} and~\ref{eval_map}), we demonstrate how our approach performs on securing the components of $Autoware$, which is the self-driving system of our \emph{ACRONIS Self-Driving Car}  (modified from a Toyota \textit{COMMS} Electrical Vehicle with a Velodyn VLP-16 Lidar, a Delphi ESR  2.5 radar, a MTi-G-710-2A8G4 GPS/IMU module and a BFS-PGE-31S4C-C camera).
In our early work, we separated $Autoware$ into 8 partitions: sensing, localization, data loading (mapping), fusion, object tracking, path planning, motion planning, and path following, where sensing, localization and mapping are isolated by partitions based on virtualization.
Experiments were conducted on real data gathered from field tests. 

In the second part (Sec.~\ref{eval_adfa}), we used the ADFA-LD data set to prove the generalization of our method.
ADFA-LD is released for host-based anomaly detection, replacing the existing benchmarks such as the KDD-98 and UNM data sets, whose applicability to modern computer systems is suspected.
Besides, ADFA-LD has a much larger degree of similarity between attack data and normal data than the KDD collections, which is more complex and harder for detection analysis~\cite{adfald}.
Thus, testing with ADFA-LD can further validate our method.

\subsection{Testing on GNSS Localization Partition}\label{eval_gnss}
In this test, we used the sample data recorded in Japan and provided by $Autoware$ project as the input of the self-driving system.
To get valid execution traces, we first ran the system and recorded all syscalls issued by GNSS localization partition (mainly the $nmea2tfpose$ program), which provided us 479 normal traces as the 1-second monitoring window was used.
Each normal trace contained 1274 syscalls averagely.
We divided those traces into the training set (240 traces for model training) and the normal testing data set (239 traces).

Then, we ran a malicious program to act as an adversary. 	
Such a program was coded to work as follows.
It stealthily gathered and sent resource usage data periodically to a remote server, and issued other critical syscalls trying to disrupt any normal execution or crash other critical services.
The whole process was to simulate a hijacked program (e.g., Trojan), which was based on a legitimate system monitor agent embedded in each partition.
After the test, an abnormal data set of 479 invalid traces (as abnormal testing data) was gathered.
Each abnormal trace' length was 1479 averagely.

We first trained the detection model with the training set where all data was normal, then tested the model with both normal and abnormal testing data.
To train the model, we first extracted features of the pattern clusters from L-$1$ to L-$15$ based on Algorithm \ref{alg:cs}.
With GNU bash $time$ command, we recorded the time overhead of the execution of our python implementation.
The total time spent in parallel extracting the features of these clusters was 6 minutes and 55.0 seconds.

\begin{table*}[!tbp]
	\centering
	\caption{Early results of the evaluation on Localization with single cluster [\%].}
	\footnotesize
	\begin{tabular}{|r|r|r|r|r|r|r|r|r|r|r|}
		\hline
		\multirow{2}{*}{\diagbox{$k$}{$\nu$}} & \multicolumn{ 2}{c|}{0.5} & \multicolumn{ 2}{c|}{0.2} & \multicolumn{ 2}{c|}{0.1} & \multicolumn{ 2}{c|}{0.05} & \multicolumn{ 2}{c|}{0.01} \\ \hline
		\multicolumn{1}{|l|}{} & \multicolumn{1}{l|}{FAR} & \multicolumn{1}{l|}{DR} & \multicolumn{1}{l|}{FAR} & \multicolumn{1}{l|}{DR} & \multicolumn{1}{l|}{FAR} & \multicolumn{1}{l|}{DR} & \multicolumn{1}{l|}{FAR} & \multicolumn{1}{l|}{DR} & \multicolumn{1}{l|}{FAR} & \multicolumn{1}{l|}{DR} \\ \hline
		1 & 50.833  & 100.000  & 19.167  & 99.791  & 12.917  & 99.582  & 5.833  & 99.374  & 0.000  & 36.326  \\ \hline
		2 & 45.417  & 99.791  & 18.333  & 99.791  & 12.500  & 99.791  & 7.083  & 99.582  & 0.417  & 36.534  \\ \hline
		3 & 50.833  & 100.000  & 27.083  & 99.791  & 13.750  & 99.791  & 6.250  & 99.791  & 2.917  & 37.161  \\ \hline
		4 & 54.167  & 100.000  & 27.500  & 100.000  & 13.750  & 100.000  & 7.083  & 100.000  & 1.250  & 99.791  \\ \hline
		5 & 61.667  & 100.000  & 33.333  & 100.000  & 24.583  & 100.000  & 16.250  & 100.000  & 8.333  & 100.000  \\ \hline
		6 & 75.833  & 100.000  & 55.417  & 100.000  & 45.417  & 100.000  & 35.833  & 100.000  & 22.083  & 100.000  \\ \hline
		7 & 89.583  & 100.000  & 80.833  & 100.000  & 75.417  & 100.000  & 65.000  & 100.000  & 43.750  & 100.000  \\ \hline
		8 & 96.250  & 100.000  & 94.167  & 100.000  & 91.667  & 100.000  & 87.083  & 100.000  & 75.000  & 100.000  \\ \hline
		9 & 99.167  & 100.000  & 96.667  & 100.000  & 95.417  & 100.000  & 94.583  & 100.000  & 92.500  & 100.000  \\ \hline
		10 & 99.167  & 100.000  & 97.917  & 100.000  & 97.500  & 100.000  & 94.167  & 100.000  & 94.583  & 100.000  \\ \hline
		11 & 99.583  & 100.000  & 99.583  & 100.000  & 99.167  & 100.000  & 98.333  & 100.000  & 99.583  & 100.000  \\ \hline
		12 & 100.000  & 100.000  & 100.000  & 100.000  & 100.000  & 100.000  & 98.750  & 100.000  & 99.167  & 100.000  \\ \hline
		13 & 100.000  & 100.000  & 100.000  & 100.000  & 100.000  & 100.000  & 99.583  & 100.000  & 99.583  & 100.000  \\ \hline
		14 & 100.000  & 100.000  & 100.000  & 100.000  & 100.000  & 100.000  & 100.000  & 100.000  & 100.000  & 100.000  \\ \hline
		15 & 100.000  & 100.000  & 100.000  & 100.000  & 100.000  & 100.000  & 100.000  & 100.000  & 100.000  & 100.000  \\ \hline
	\end{tabular}
	\label{av_singles}
\end{table*}

As shown in Table~\ref{av_singles}, after applying the method proposed in Sec.~\ref{training}, the evaluation results indicated that
L-$1$ to L-$8$ clusters 
could form a potential candidate set as they provided better performance than others when $\nu$ = 0.01.

\begin{figure}[tbp]
	\centering
	\includegraphics[width=0.7\linewidth]{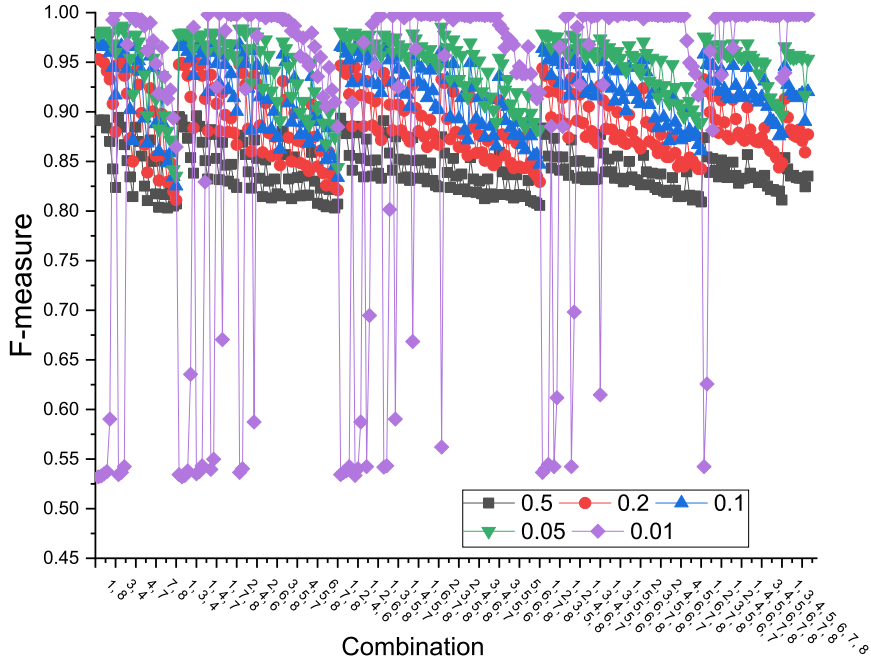}
	\caption{F-measure of the evaluations on localization partition with different $\nu$.}
	\label{fm_awlocal}
	\vspace{-10pt}
\end{figure}

\begin{figure}[t!]
	\centering
	\includegraphics[width=0.7\linewidth]{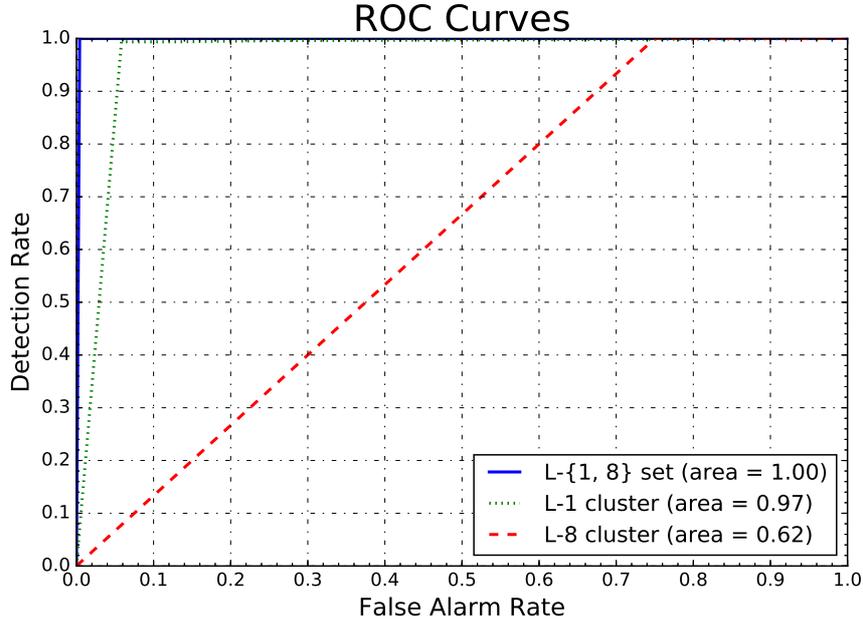}
	\caption{ROC curves for the evaluations of the execution of localization partition ($AREA=AUC$).}
	\label{roc_nmea}
\end{figure}

Then, we used the extracted feature data sets in the previous evaluation to generate the inputs of SVM under various combinations of different clusters.
The total number of combinations is $2^8 - (8+1) = 247$, which took 8.2s (recorded by $time$ command) to train and test them in parallel.
Among all trials, we found that when $\nu = 0.01$, combinations such as \{1, 8\}, \{1, 4, 8\}, \{1, 5, 8\}, \{1, 6, 8\}, \{1, 3, 5, 8\}, \{1, 3, 6, 8\}, \{1, 3, 7, 8\}, \{1, 4, 5, 8\}, \{1, 4, 6, 8\}, \{1, 4, 7, 8\} and \{1, 5, 6, 8\}, etc. offered the best performance (F-measure = 0.999), as shown in Fig.~\ref{fm_awlocal}.
Note that 10-fold cross validation was adopted to reduce over-fitting during the training for each combination.
We chose $\left\lbrace 1, 8 \right\rbrace$ since it was the shortest combination.
The ROC (Receiver Operating Characteristic) curves of different combinations were also shown in Fig.~\ref{roc_nmea}, which proved that using the combination of
L-$1$ and L-$8$ clusters, i.e., \{1, 8\}, significantly improved the classification performance. 

We compared our approach with single-cluster classifiers and a multi-voter classifier.
A single-cluster classifier is a one-class SVM trained by the feature of a single cluster, and the multi-voter classifier contained several single-cluster classifiers.
The voter collected results from those individual classifiers and returned the final result with their sum.
If the sum is negative, then the voter returns ``abnormal'', otherwise ``normal''.
The results were shown in Table~\ref{nmea_com}.
We can conclude that compared with single-cluster classifiers, multi-cluster classifier decreased FAR and maintained high DR, while the multi-voter classifier performed the worst by providing the highest FAR.

\begin{table}[t!]
	\caption{Comparison on Localization Partition}
	\footnotesize
	\begin{center}
		\begin{tabular}{|c|r|r|}
			\hline
			Classifier & \multicolumn{1}{c|}{FAR(\%)} & \multicolumn{1}{c|}{DR(\%)} \\ \hline
			\{1, 8\}  & 0.417 & 100.000 \\ \hline
			\{1\}  & 5.833 & 99.374 \\ \hline
			\{8\} & 75.000 & 100.000 \\ \hline
			\{1\} $+$ \{8\} & 73.000 & 100.000 \\ \hline
		\end{tabular}
	\end{center}
	\label{nmea_com}
\end{table}

\subsection{Testing on Mapping Partition}\label{eval_map}
\begin{figure}[t!]
	\centering
  \subfigure[Map of the testing car park area.]{  \label{sing_map}
      \begin{minipage}[b]{0.42\linewidth}
         \centering
         \includegraphics[width=\linewidth]{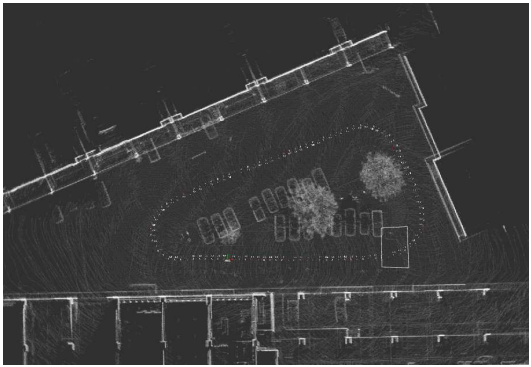}
      \end{minipage}}
      \hspace{3pt}
  \subfigure[Our Toyota COMS AV.]{\label{car}
      \begin{minipage}[b]{0.38\linewidth}
         \centering
         \includegraphics[height=0.8\linewidth]{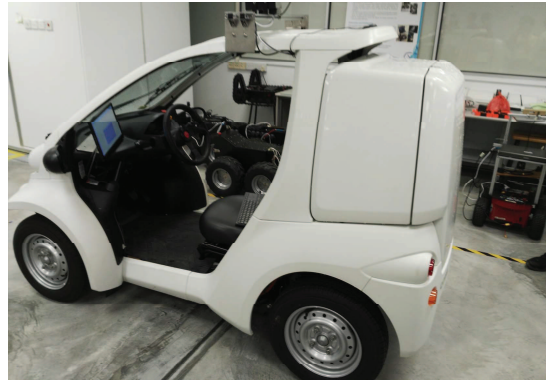}
      \end{minipage}}
  \caption{Our \av and the map generated by the car in the testing car park.}
\end{figure}

During the test, we used the mapping data gathered from our field test in Singapore as the workload for the target software. 
The mapping partition broadcast vector maps and point-cloud maps of the driving area, as shown in Fig.~\ref{sing_map}, which were collocated by our self-driving car in a field test, as shown in Fig.~\ref{car}.
Our self-driving car circled the area following those waypoints in the map.
Meanwhile, all syscalls issued by the Mapping partition were captured and recorded.
Like in Sec.~\ref{eval_gnss}, the model was built with training data (normal traces), and evaluated with both normal and abnormal testing data.

Like the previous setup, we collected a normal data set consisting of 300 traces with 1-second monitoring window.
Those sequences were divided evenly into model training and normal testing data.
After another run with the aforementioned malicious program embedded, 300 abnormal traces were gathered in 5 minutes, which were used as abnormal testing data.
Among those data, each normal trace contained 617 elements, while each abnormal one had 724, averagely.

\begin{table*}[htbp]
	\caption{Early results of using single pattern set on Mapping [\%].}
	\footnotesize
	\centering
	\begin{tabular}{|l|r|r|r|r|r|r|r|r|r|r|}
		\hline
		& \multicolumn{ 2}{c|}{0.5} & \multicolumn{ 2}{c|}{0.2} & \multicolumn{ 2}{c|}{0.1} & \multicolumn{ 2}{c|}{0.05} & \multicolumn{ 2}{c|}{0.01} \\ \hline
		& \multicolumn{1}{l|}{FAR} & \multicolumn{1}{l|}{DR} & \multicolumn{1}{l|}{FAR} & \multicolumn{1}{l|}{DR} & \multicolumn{1}{l|}{FAR} & \multicolumn{1}{l|}{DR} & \multicolumn{1}{l|}{FAR} & \multicolumn{1}{l|}{DR} & \multicolumn{1}{l|}{FAR} & \multicolumn{1}{l|}{DR} \\ \hline
		1 & 56.000  & 100.000  & 17.333  & 99.667  & 10.000  & 99.667  & 2.667  & 95.333  & 0.667  & 93.667  \\ \hline
		2 & 33.333  & 99.667  & 14.000  & 99.667  & 10.000  & 99.333  & 2.000  & 98.000  & 0.000  & 97.333  \\ \hline
		3 & 48.667  & 100.000  & 20.667  & 100.000  & 16.667  & 100.000  & 8.667  & 100.000  & 6.000  & 100.000  \\ \hline
		4 & 52.667  & 100.000  & 24.000  & 100.000  & 16.000  & 100.000  & 14.000  & 100.000  & 7.333  & 100.000  \\ \hline
		5 & 65.333  & 100.000  & 40.000  & 100.000  & 27.333  & 100.000  & 22.000  & 100.000  & 12.667  & 100.000  \\ \hline
		6 & 87.333  & 100.000  & 73.333  & 100.000  & 63.333  & 100.000  & 47.333  & 100.000  & 36.667  & 100.000  \\ \hline
		7 & 93.333  & 100.000  & 86.000  & 100.000  & 83.333  & 100.000  & 80.667  & 100.000  & 62.000  & 100.000  \\ \hline
		8 & 99.333  & 100.000  & 98.667  & 100.000  & 98.000  & 100.000  & 97.333  & 100.000  & 93.333  & 100.000  \\ \hline
		9 & 99.333  & 100.000  & 99.333  & 100.000  & 99.333  & 100.000  & 98.667  & 100.000  & 98.667  & 100.000  \\ \hline
		10 & 100.000  & 100.000  & 100.000  & 100.000  & 100.000  & 100.000  & 100.000  & 100.000  & 100.000  & 100.000  \\ \hline
		11 & 100.000  & 100.000  & 100.000  & 100.000  & 100.000  & 100.000  & 100.000  & 100.000  & 100.000  & 100.000  \\ \hline
		12 & 100.000  & 100.000  & 100.000  & 100.000  & 100.000  & 100.000  & 100.000  & 100.000  & 100.000  & 100.000  \\ \hline
	\end{tabular}
	\label{mapping_com}
\end{table*}

\begin{figure}[t!]
	\centering
	\includegraphics[width=0.7\linewidth]{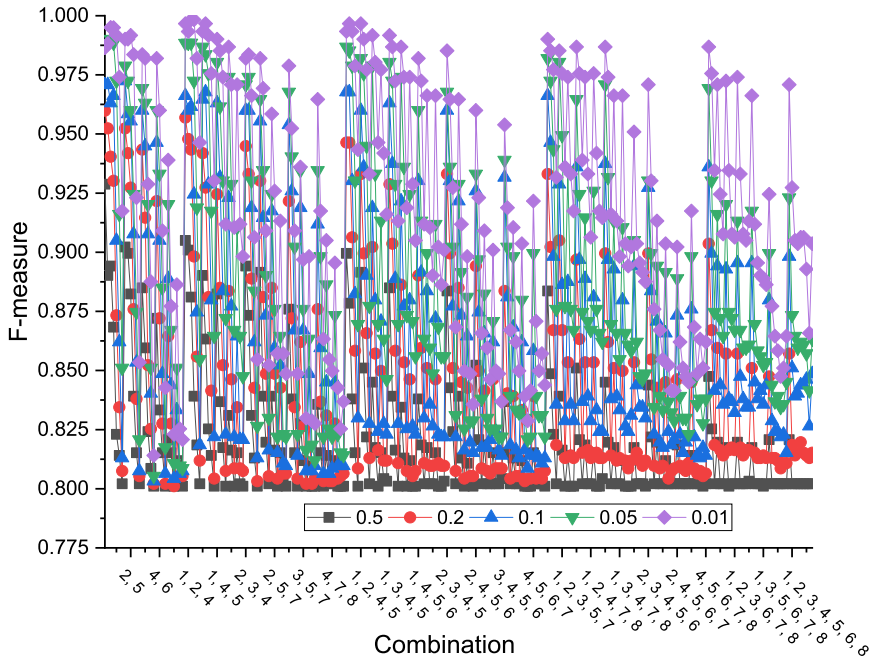}
	\caption{F-measure of the evaluations on mapping partition with different $\nu$.}
	\label{fm_map}
	\vspace{-10pt}
\end{figure}

\begin{figure}[t!]
	\centering
	\includegraphics[width=0.7\linewidth]{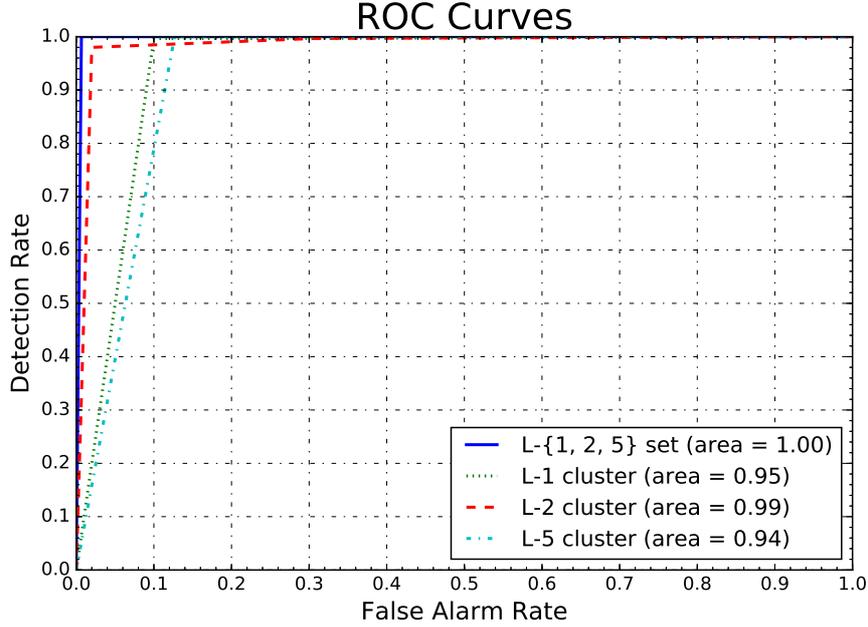}
	\caption{ROC curves for the evaluations of the execution of mapping partition ($AREA=AUC$).}
	\label{roc_map}
\end{figure}

The total time for parallel feature extraction of L-1 to L-12 clusters took 1 minute and 29.0 seconds.
Among the results shown in Table~\ref{mapping_com}, L-1 to L-8 clusters can form the pattern set as $\Delta_{k,0.01}$ reaches the minimum value when $k=8$.
Thus, the number of possible combinations was $2^8 - (8+1) = 247$, which took 6.31s to test them using the grid searching.
The F-measure results are shown in Fig.~\ref{fm_map}, from which we could conclude that when $\nu$ = 0.01, each combination achieved the best performance.
Among the achieved results, we found combinations of
\{1, 2, 5\} and \{1, 2, 6\} 
yielded better results, achieving the best F-measure 0.998.
Thus, \{1, 2, 5\} was chosen as it was the shortest combination, as shown in Fig.~\ref{roc_map}.
The comparison is shown in Table~\ref{map_com}, which suggested that the multi-cluster classifier achieved better results than single-cluster ones.

\begin{table}[t!]
	\caption{Comparison on Mapping partition.}
	\footnotesize
	\begin{center}
		\begin{tabular}{|c|r|r|}
			\hline
			Classifier & \multicolumn{1}{c|}{FAR(\%)} & \multicolumn{1}{c|}{DR(\%)} \\ \hline
			\{1, 2, 5\} & 0.667 & 100.00 \\ \hline
			\{1\} & 10.000 & 99.667 \\ \hline
			\{2\} & 0.000 & 97.333  \\ \hline
			\{5\} & 12.667 & 100.000  \\ \hline
			\{1\} $+$ \{2\} $+$ \{5\} & 8.00 & 100.00 \\ \hline
		\end{tabular}
	\end{center}
	\label{map_com}
\end{table}

From the above two tests, we could find that combining multiple clusters did improve detection accuracy.
In the first simulation, combining \{1, 8\} reduced FAR by 92.85\% (compared with only using L-1 cluster), and achieved 100\% DR.
In the second one, \{1, 2, 5\} reduced FAR by 93.33\% (compared with using L-1 cluster only), and achieved 100\% DR.
Even when compared with the multi-voter classifier, it greatly reduced FAR by over 90.0\%.
The comparison showed that by combining multiple features, it can achieve the minimum FAR and maximum DR among all single sets used.

\subsection{Comparison with Other Methods on ADFA-LD}\label{eval_adfa}
In both the above cases, most self-driving programs were running periodically, and the control logic was not very complex, which might lead to the uniformity of syscalls and raise the similarity concern, i.e. normal/abnormal data distributed differently to make detection easier.
Although we had shown that it was hard to distinguish the data (Table~\ref{av_singles} and~\ref{mapping_com}) with single features (similar to methods in~\cite{mxie2}), we would further evaluate our approach on the public ADFA-LD data set and compare the result with other similar methods.

The ADFA-LD contains 833 training traces, 4372 validation (normal) traces and 746 attack (abnormal) traces.
Each training trace contains about 370 syscalls, each validation (normal testing) trace has 485 items, and each attack (abnormal) trace has 426 elements, averagely.
In the test, we used the training set to tune the classifier, and tested it on the validation and attack sets.

\begin{figure}[t!]
	\centering
	\includegraphics[width=0.7\linewidth]{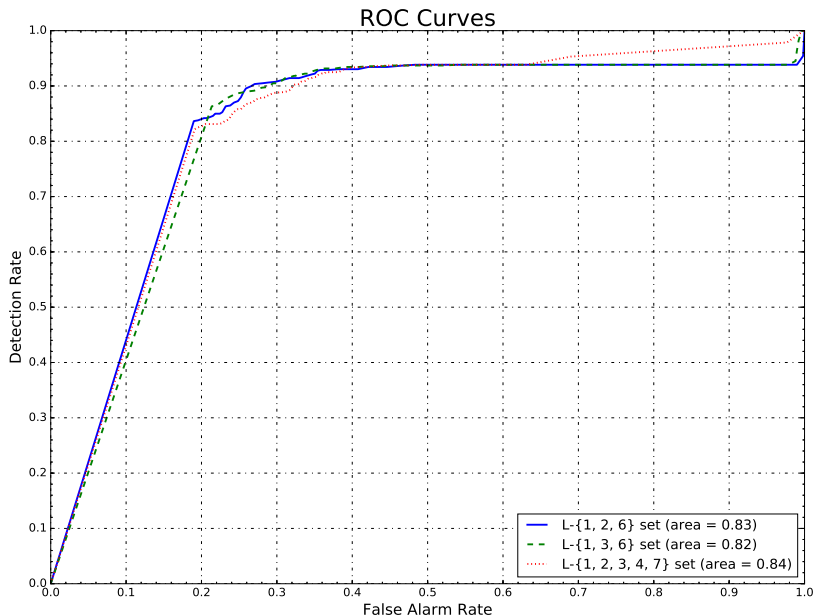}
	\caption{ROC curves of evaluations on ADFA-LD ($AREA=AUC$).}
	\label{roc_adfa}
\end{figure}

\begin{figure}[t!]
	\centering
	\includegraphics[width=0.7\linewidth]{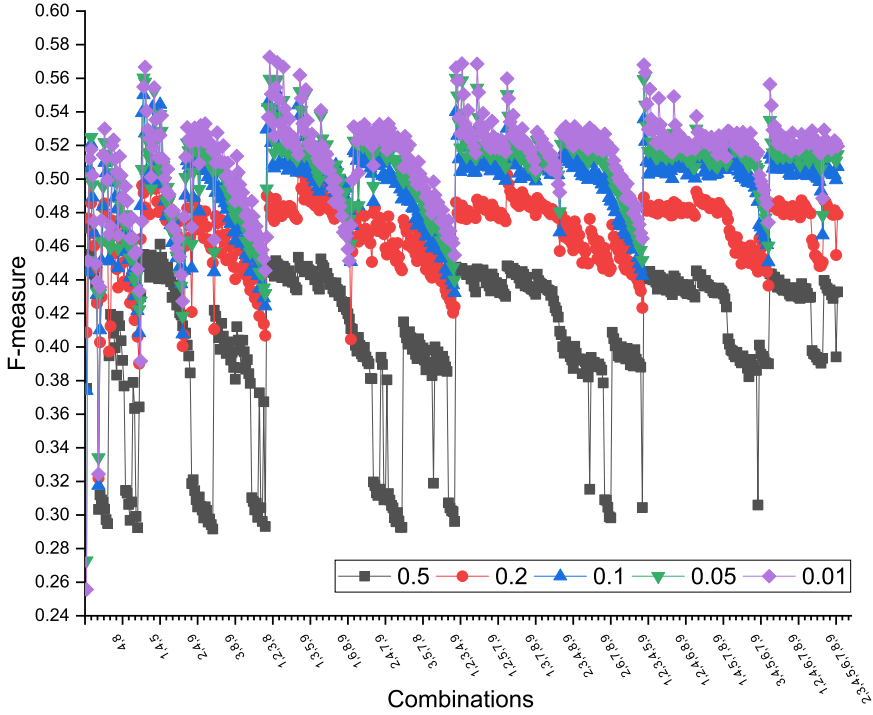}
	\caption{F-measure of evaluations on ADFA-LD with different $\nu$.}
	\label{fm_adfa}
\end{figure}

From the early results in Table~\ref{adfa-single}, we set $\nu=0.01$ and used L-1 to L-9 clusters to find a proper combination.
With GNU bash $time$ command, the recorded time for parallel feature extraction was 24 minutes and 7 seconds, including building the clusters and computing their frequencies.
Training and testing $2^9 - (9+1) = 502$ trials took 1 minute and 19.8 second.
Among all attempts, we found combinations \{1, 2, 6\}, \{1, 3, 6\}, \{1, 2, 5, 6\}, \{1, 3, 4, 6\}, \{1, 2, 3, 4, 7\} and \{1, 2, 3, 5, 6\} had better performance (higher F-measures), as shown in Fig. \ref{fm_adfa}.
However, \{1, 2, 5, 6\}, \{1, 3, 4, 6\} and \{1, 2, 3, 5, 6\} were removed since they can be regarded as extensions of \{1, 2, 6\} and \{1, 3, 6\}.
We compared \{1, 2, 6\}, \{1, 3, 6\} and \{1, 2, 3, 4, 7\} and showed their ROC curves in Fig.~\ref{roc_adfa}.
Even though their AUCs (Area Under the Curve) were almost the same, \{1, 2, 6\} provided better classification performance than others since it has the largest F-measure with $DR = 83.6\%$ and $FAR = 18.9\%$.

We compared the achieved results with ones reported by the following works, among which either the SVM based solution or other fast detection method was chosen, and the same amount of data was used for training and testing. Those included:  
Creech \emph{et al}~\cite{creech} used one-class SVM with semantic features, achieving DR of 80\% at 15\% FAR, but the training overhead was expensive as it took weeks for extracting the semantic features.
Marteau~\cite{tifs19} used covering similarity with only the training data set (no additional training data from the validation set), achieving DR of 80\% at 19\% FAR.
Xie \emph{et al}~\cite{mxie1-1,mxie1-2} combined k-nearest neighbors (k-NN) and k-mean clustering approaches and achieved DR of 60\% at FAR of 20\%,
In another work~\cite{mxie2}, they used short sequences and frequency features to train a one-class SVM, and achieved DR of 70\% at FAR of 20\%.
Haider \emph{et al}~\cite{whaider} used four statistical features, i.e., the least/most repeated and the minimum/maximum values in a trace, to represent a trace and detect attacks.
They used three supervised learning algorithms: SVMs with linear and radial basis kernels and k-NN, and the best result was k-NN with a 78\% DR at 21\% FAR.
As shown in Table~\ref{adfa_com}, compared with those best results quoted from each publication, our proposed method could get a good detection performance while still took only minutes' training.
For example, compared with~\cite{tifs19}, our approach achieves better results while reducing the size of the pattern set with the proposed method in Sec.~\ref{three-steps}, and our approach runs much faster than~\cite{creech} while maintaining similar performance.
Please note that some papers only reported the feature extraction time (the overall execution time will be longer).
\begin{table}[t!]
	\caption{Comparison of different methods on ADFA-LD.}
	\footnotesize
	\begin{center}
		\begin{tabular}{|m{3cm}|m{1cm}|m{1cm}|m{1.61cm}|}
			\hline
			Methods & FAR(\%) & DR(\%) & Training Time \\ \hline
			Marteau~\cite{tifs19}, training the SVM with only the training data set & 19.0 & 80.0 & minutes \\ \hline
			Creech \emph{et al}~\cite{creech}: SVM with semantic features  & 15.0 & 80.0 & weeks \\ \hline
			 Xie \emph{et al}~\cite{mxie1-1,mxie1-2}: k-NN and k-mean clustering & 20.0 & 60.0 & seconds\\ \hline
			Xie \emph{et al}~\cite{mxie2}: frequencies of short sequences and one-class SVM  & 20.0 & 70.0 & seconds \\ \hline
			Haider \emph{et al}~\cite{whaider}: statistical features and SVM & 21.0 & 78.0 & seconds \\
			\hline
			\textbf{Proposed method} & \textbf{18.9} & \textbf{83.6} & \textbf{minutes} \\ \hline
		\end{tabular}
	\end{center}
	\label{adfa_com}
	\vspace{-15pt}	
\end{table}

\vspace{-5pt}
\subsection{Overhead analysis}
The achieved results show that feature extraction takes more time than training SVM models.
The process of feature extraction traverses the entire training data set and calculates the feature vector of each trace in both training and testing data sets.
Hence, such a process is subjective to the data amount.
However, since each extraction attempt is independent, such a process can run in parallel to extract a single L-$k$ cluster and calculate the frequency features.
Hence, only the maximum extracting time matters.
The extraction is essentially a string search process, and the number of traces $n$, the maximal trace length $l$ and the number of desired clusters $m$ affect the processing time greatly.
The time complexity of such a process is $O(nlm)$.
In our evaluation, training a one-class SVM doesn't take much time.
Although the input data size matters, training an SVM model takes averagely less than a second in our experiments.

%% file: chapter6.tex

\section{Conclusion and Future Work}\label{conclusion}
In this paper, we propose an inexpensive detection method using syscalls to monitor the execution of critical software functions of self-driving systems.
Such a method extracts the syscall pattern of normal traces and uses the invocation frequency of multiple pattern clusters as featured inputs to train a one-class SVM based detection model.
The evaluation shows the proposed method can further reduce the false alarm rate and maintain high accuracy based on the test on real data obtained from the self-driving system \emph{Autoware}.
A further comparison against existing works with the ADFA-LD data set demonstrates that such a method improves the detection performance with short training time.

This work shows that combining multiple features could improve detection performance. 
It is also possible to apply such a method in other domains.
In the future, we plan to extend the approach by stealthily intercepting other critical APIs to trace programs' behaviors (e.g. computation) more accurately, or by mapping several contiguous syscalls to an API invocation (e.g. $ros::Time::now()$ and its corresponding set of syscalls).  
In such a way, we may better understand the execution, which could help model the self-driving functions more precisely.